\documentclass[twocolumn,aps,prl,showpacs,groupedaddress]{revtex4}
\usepackage{epsfig,amssymb,amsmath}
\usepackage[hypertex,linkcolor=red]{hyperref}

\def\section#1{{\em #1: ---}}
\def\comment#1{}\def\labell#1{\label{#1}}
\def\togli#1{}

\newcommand{\ket}[1]{|#1\rangle}

\begin{document}

%\fbox{{\scriptsize Submitted paper.}}
%Title of paper
\title{Experimental Quantum Private Queries with linear optics}

\author{Francesco De Martini$^{1,2}$, Vittorio Giovannetti$^3$, Seth
  Lloyd$^4$, Lorenzo Maccone$^5$, Eleonora Nagali$^1$, Linda
  Sansoni$^1$, and Fabio Sciarrino$^1$}\affiliation{$^1$ Dipartimento
  di Fisica dell'Universit\`{a} ``La Sapienza'',
  Roma 00185, Italy\\$^2$ Accademia Nazionale dei Lincei, via della Lungara 10, Roma 00165,
Italy
   \\$^{3}$ NEST-CNR-INFM \& Scuola Normale
  Superiore, Piazza dei Cavalieri 7, 56126, Pisa, Italy.\\$^{4}$ MIT,
  RLE and Dept. of Mech. Engin. MIT 3-160, 77 Mass.~Av., Cambridge, MA
  02139, USA.\\ $^5$Institute for Scientific Interchange, 10133
  Torino, and QUIT, Dip.~A.~Volta, 27100 Pavia, Italy.}
\begin{abstract}
  The Quantum Private Query is a quantum cryptographic protocol to
  recover information from a database, preserving both user and data
  privacy: the user can test whether someone has retained information
  on which query was asked, and the database provider can test the
  quantity of information released. Here we introduce a new  variant Quantum Private Query algorithm  
which  admits a simple linear optical implementation: it employs the photon's momentum (or time slot) as
  address qubits and its polarization as bus qubit.  A
  proof-of-principle experimental realization is implemented.
\end{abstract}
\pacs{03.67.-a,03.67.Lx,03.67.Dd}
%03.67.-a Quantum information
%03.67.Dd Quantum cryptography
%03.67.Lx Quantum computation
%03.67.Mn Entanglement production, characterization, and manipulation  (see also 03.65.Ud Entanglement and quantum nonlocality; for entanglement in Bose-Einstein condensates, see 03.75.Gg)

\maketitle

Quantum information technology has matured especially in the field of
cryptography\togli{~\cite{Benn84}}. Two distant parties can exploit
quantum effects, such as entanglement, to communicate in a provably
secure fashion\togli{: the very laws of physics prevent a third party
  to intercept their communication. This field is already moving from
  basic research to technological implementations, as various
  companies are thriving by producing commercial quantum-cryptographic
  products. }. An interesting cryptographic primitive is the
Symmetrically-Private Information Retrieval (SPIR)~\cite{Gert00}: it
allows a user (say Alice) to recover an element from a database in
possession of a provider (say Bob), without revealing which element
was recovered ({\em user privacy}). At the same time it allows Bob to
limit the total amount of information that Alice receives ({\em data
  privacy}). Since user and data privacy appear to be conflicting
requirements, all existing classical protocols rely on constraining
the resources accessible by the two parties~\cite{amba}. However,
using quantum effects, such constraints can be dropped: the Quantum
Private Query (QPQ)~\cite{qpq} is a quantum-cryptographic protocol
that implements a cheat-sensitive SPIR.  User privacy is indirectly
enforced by allowing Alice to test the honesty of Bob: she can perform
a quantum test to find out whether he is retaining any information on
her queries, in which case Bob would disturb the states Alice is
transmitting\togli{, analogously to what happens in conventional
  quantum cryptography~\cite{Benn84}), but here a cheating Bob has
  some probability of evading detection~\cite{qpq,security}.} and she
has some probability of detecting it~\cite{security}. Data privacy is
strictly enforced since the number of bits that Alice and Bob exchange
is too small to convey more than at most two database items.

In this paper we present an optical scheme to carry out a variant of
the QPQ protocol. In contrast to the original proposal of~\cite{qpq},
it does not require a quantum random access memory (qRAM)~\cite{qram}
and can be implemented with linear optics, i.e.~current technology,
but it has sub-optimal communication complexity. The qRAM's absence
implies that the binary-to-unary translation to route Alice's query to
the appropriate database memory element must be performed by Alice
herself. Thus Alice and Bob must be connected by a number of
communication channels equal to the number $N$ of database elements
(although $O(\log N)$ would suffice with a qRAM).  We present two
conceptually equivalent QPQ implementations: in the first (more suited to explanatory purposes and proof-of-principle tests)
each channel is a spatial optical mode, in the
second (more suited to practical applications) it is a time slot in a
fiber~\cite{gisin1,gisin2}. 
The paper focuses mostly on  the former implementation for which we provide  an experimental test. 
For this setup we 
also consider  the case in which 
Alice entangles her queries with ancillary systems that she keeps in her lab. With this choice  the  user privacy  
can only be enhanced with respect to original scheme~\cite{qpq}  as
Bob has only limited access to the 
states which encode Alice's queries.

We start with a description of the new scheme, focusing on how user
and data privacy can be tested. Then we describe its experimental
implementation, and conclude with the time-slot implementation.

%%%%%%%%%%%%%%%%%%%%%%%%%%%%%%%%%%%%%
%\begin{widetext}
\begin{figure}[t]
\begin{center}
\includegraphics[width=7cm]{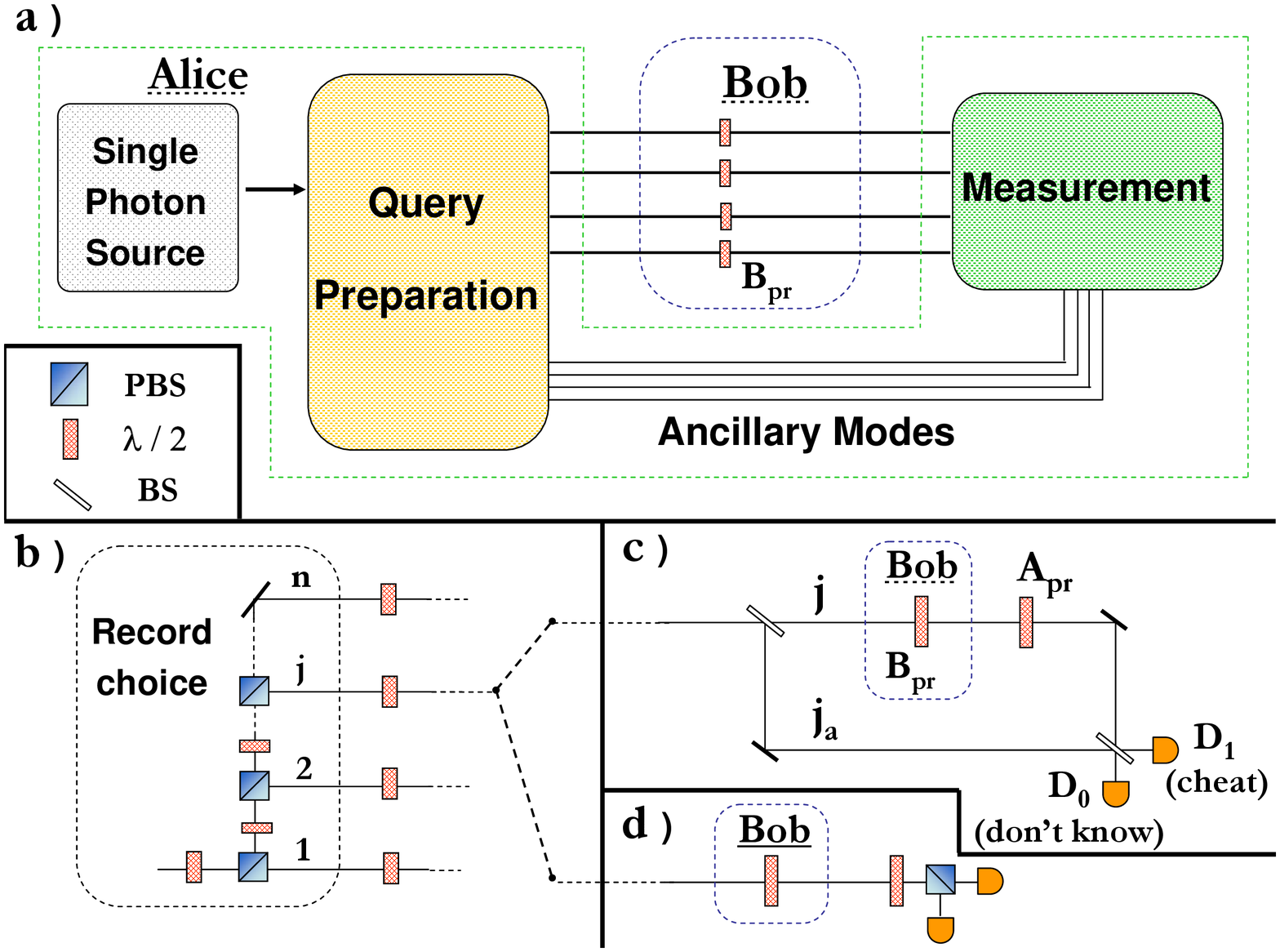}
\end{center}
%\vspace{-.5cm}
\caption{\textbf{a)}~Overview of the experiment.  Alice at the
  query-preparation stage routes a single photon to the appropriate
  spatial modes, where Bob's database are stored in an array of
  polarization rotators B$_{pr}$. 
  \textbf{b)}~Alice's query preparation stage. A set of half-wave
  plates and polarizing beam splitters route the photon into the
  spatial mode $j$ chosen by Alice.  She also chooses whether to send
  a ``superposed'' query (see~\textbf{c}) or a ``plain'' query
  (see~\textbf{d}).  \textbf{c)}~If she chose a superimposed query
  $\ket{{S}_j}$, Alice performs the honesty test through an
  interference experiment in the $j$-th mode. 
  \textbf{d)}~Instead, if she chose a plain query $|{P}_j \rangle$,
  she performs a polarization measurement on the photon to recover the value of $A_{j}$.}
\labell{f:schemablocchi}
\end{figure}
%\end{widetext}
%%%%%%%%%%%%%%%%%%%%%%%%%%%%%%%%%%%%%%%%%%%%%

\textbf{The scheme. }The optical QPQ scheme is sketched in
Fig.~\ref{f:schemablocchi}(a).  Bob controls an $N$-element database,
where each element $j$ is associated to a spatial optical mode and
consists of one bit $A_j$ of classical information. The bit $A_j=1$
$(0)$ is encoded into the presence (absence) of a half-wave plate
B$_{pr}$ in the $j$th mode (it rotates the polarization by
$90^\circ$).  Alice probes this system with single photons either in
one mode or in a superposition of modes. To recover the database
element $A_j$, Alice sends to Bob a single horizontally polarized
photon $H$ in the mode $j$, i.e.~the state $\ket{{P}_j }= \ket{H}_j$--
see Fig.~\ref{f:schemablocchi}(b)\togli{: she uses the photon's
  spatial mode $j$ to encode the ``address'' of the database entry she
  wants to access}.  Bob employs the photon's polarization as a
``bus'' qubit to communicate the query result: vertical $V$ if
$A_j=1$, or horizontal $H$ if $A_j=0$. Namely, his transformation is
\begin{eqnarray} \label{new1} \ket{{P}_j}\ \to\ \ket{{{P}}_j^{out} }=
  \ket{A_j}_j\equiv\left\{\begin{matrix}\ket{H}_j&\mbox{ for
      }&A_j=0&\cr\ket{V}_j&\mbox{ for }&A_j=1&\;.\end{matrix}\right.
\end{eqnarray}
This exchange is clearly not private\togli{ since Bob can easily read
  Alice's question without her knowledge (he has simply to detect in
  which spatial mode Alice is sending her photon, and then re-create
  an identical photon in the same mode)}.  To attain cheat
sensitivity, Alice must randomly alternate two different kinds of
queries: ``plain'' queries of the type $\ket{{P}_j}$ described above
and ``superposed'' queries
\begin{eqnarray}
  \ket{{S}_j }=
  (\ket{H}_j\ket{\varnothing}_{j_a}+|\varnothing\rangle_{j}
|H\rangle_{j_a})/\sqrt{2}
\labell{quer}\;,
\end{eqnarray}
where the $j$-th mode is entangled with an ancillary spatial mode
$j_a$, and where $|\varnothing\rangle$ is the vacuum
state. 
According to original proposal~\cite{qpq}  $j_a$ should
 be identified with one  (say the first) of the $N$ spatial modes of the system, 
whose associated database entry is initialized in a known fiduciary value $A_{j_a}=0$.
 With this choice $j_a$ will play the role of the {\em rhetoric}
 query of the original QPQ scheme
 whose user privacy has been formally proved in Ref.~\cite{security}. 
Here however we  follow an alternative strategy which guarantees user privacy levels which are at least
as good as the original scheme and which can be easily realized in the spatial mode implementation.
Namely, as shown in Fig.~\ref{f:schemablocchi}(a), 
the $j_a$'s will be identified 
with extra modes that Alice keeps in her lab. With this choice  
Alice privacy  can only increase with respect to the original scheme as
Bob does not have access to the complete quantum system~(\ref{quer}) --- his cheating
operations can only act on the subsystem that Alice has sent him while  the 
QPQ security proof~\cite{security} assumes he can act on the full state system.
To prepare such  input state Alice simply shines an $H$
polarized photon onto a $50\%$ beam-splitter sending one of the emerging beams to Bob and keeping the other in her lab as shown in
Fig.~\ref{f:schemablocchi}(c). After having crossed Bob's lab (in the
absence of cheating), the superposed query is evolved into
\begin{eqnarray}
\ket{{S}_j}\ \to\  \ket{{{S}}_j^{out} } =  (|A_j\rangle_j |\varnothing\rangle_{j_a}
  +|\varnothing\rangle_{j}|H\rangle_{j_a}
  )/\sqrt{2} \labell{quer2}\;.
\end{eqnarray}
The two types of queries $\ket{{P}_j }$ and $\ket{{S}_j}$ must be
submitted {\em in random order} and {\em one at a time} (i.e.~she must
wait for Bob's first reply before sending him the second query): if
Bob received both queries at the same time, he could cheat undetected
with a joint measurement~\cite{qpq}.

The random alternation of plain and superposed queries allows Alice to
test Bob's honesty. Indeed, since he does not know whether her photon
is in the state $\ket{{P}_j}$ or $\ket{{S}_j}$, if Bob measures its
position he risks collapsing the superposed query $\ket{{S}_j}$, and
Alice can easily find it out.  In fact, she can first obtain the value
of $A_j$ through a polarization readout from $|{{P}}_j^{out}\rangle$
-- see Fig.~\ref{f:schemablocchi}(d). She can then use this value to
prepare a projective measurement that tests whether the superposed
query $\ket{{S}_j}$ has been preserved or collapsed (honesty test),
i.e.~a measurement that tests if the answer associated with
$\ket{{S}_j}$ has been collapsed into the subspace orthogonal to the
expected output $\ket{{S}_j^{out}}$. [As explained in more detail in
the next section, this essentially amounts to the interferometric
measurement of Fig.~\ref{f:schemablocchi}(c).]  If this happened, she
can confidently conclude that Bob has cheated.  If this has not
happened she cannot conclude anything: a cheating Bob still has some
probability of passing the test.  For instance, assume that Bob uses a
measure-and-reprepare strategy on one of the two queries, he will be
caught only with probability $1/4$.  Anyhow, whatever
cheating strategy Bob may employ, the probability of passing the
honesty test is bounded by the information he retains on Alice's
query~\cite{security}: he can pass the test with certainty {\em if and
  only if} he does not retain {\em any} information from her.

\textbf{Readout and honesty test.} Before proceeding, we analyze in
more detail Alice's measurements. \togli{Alice first has to extract
  the value $A_j$ from the plain output $\ket{{P}_j^{out}}$. She then
  uses $A_j$ to prepare a projective measurement to discriminate
  $\ket{{{S}}_j^{out}}$ from its orthogonal complement, namely the
  honesty test.}  Consider first the case in which Alice {\em first}
sends the plain query $\ket{{P}_j }$ and {\em then} the superposed
query $\ket{{S}_j }$. In this case, she recovers $A_j$ with the
polarization measurement of Fig.~\ref{f:schemablocchi}(d). Then,
before sending the second query $\ket{{{S}}_j }$, she sets up an
interferometer which couples the ancillary mode $j_a$ with the output
of the mode $j$ as shown in Fig.1(c), where the polarization rotator
A$_{pr}$ is used to compensate the rotation induced by Bob's database,
determined by the value of $A_j$ that she previously recovered.
Therefore, if Bob has not cheated, the state in the interferometer
just before the second beam splitter is $\ket{{{S}}_j}$
so that the ``don't know'' detector D$_0$ {\em must} fire and the
``cheat'' detector D$_1$ {\em cannot} fire.  If the ``cheat'' detector
D$_1$ does fire, Alice knows that Bob must have cheated.

Consider now the case in which Alice sends {\em first} the superposed
query $\ket{{S}_j}$ and {\em then} the plain query $\ket{{P}_j}$. In
order to perform the honesty test, she must first recover the value of
$A_j$. So she needs to store the answer to the superposed query
$\ket{{S}_j^{out}}$ until the answer to the plain query
$\ket{{P}_j^{out}}$ arrives, from which $A_j$ can be measured. It
requires a quantum memory~\cite{Chuu08Chen08Yuan08} and a fast
feed-forward mechanism~\cite{Scia06} to prepare the honesty test
measurement depending on the value of $A_j$.  Achieving this is
possible, but demanding.  The same goal is reached with a less
efficient but much simpler strategy.  Alice chooses a random value
$A^{(R)}$ in place of $A_j$.  She then performs the interferometric
measurement of Fig.~\ref{f:schemablocchi}(c) inserting or not the
polarization rotator $A_{pr}$ depending on the value of $A^{(R)}$.
This interferometer is then a projector on the state
$|S^{(guess)}_j\rangle\equiv(|A^{(R)} \rangle_j
|\varnothing\rangle_{j_a}+|\varnothing\rangle_{j}|H\rangle_{j_a}
)/\sqrt{2}$.  Later, when she receives the output of the plain query
$\ket{{P}_j^{out}}$, she finds out the value of $A_j$. If she had
picked the right value $A^{(R)}=A_j$, she will know that her first
measurement was a valid honesty test since
$\ket{{S}^{out}_j}=\ket{{S}^{(guess)}_j}$. Otherwise, if \togli{she
  had picked the wrong value} $A^{(R)}\neq A_j$, then the result of
her honesty test is useless and she must discard it. Since Alice
chooses $A^{(R)}=A_j$ with probability $1/2$, she performs the honesty
test only on half of the transactions. This reduces her probability of
discovering a cheating Bob, but not by a huge amount. For instance, in
the example analyzed above, the probability is reduced from $1/4$ to
$3/16$\togli{~\cite{prob2}}.  As before, Bob passes the honesty test
with probability 1 if and only if he does not cheat.

Let us now briefly summarize the protocol. 1) Alice randomly chooses
one of the two scenarios: either send first the plain query
$\ket{P_j}$ and then the superposed query $\ket{S_j}$, or
viceversa. 2a) In the first case, she recovers $A_j$ from Bob's first
reply and uses it to prepare the honesty test to use on his second
reply. 2b) In the second case, she chooses a random bit $A^{(R)}$ and
prepares the honesty test using it in place of $A_j$. Then she performs
the honesty test on Bob's first reply. When $A_j$ becomes available
later (from Bob's second reply), she finds out whether the honesty test
result was meaningful (if $A_j=A^{(R)}$) or not (if $A_j\neq
A^{(R)}$). 3) If the honesty test was meaningful and it has failed, she can
conclude that Bob has cheated.

\textbf{Data privacy.} In the original QPQ protocol~\cite{qpq}, data
privacy was ensured by the fact that only a limited number of qubits
were exchanged between Alice and Bob: \togli{Alice did not have access
  to the database memory cells, but} she had to send (and receive) a
sequence of ${\cal O}(\log N)$ qubits to specify the address of the
$j$ th element.  In contrast, in this version of the protocol Alice
has direct access to all the entries of Bob's database through the $N$
optical modes. She can then violate data privacy and recover multiple
elements of Bob's database by sending many photons, one per mode.
Theoretically, Bob can foil Alice by performing a joint measurement on
the $N$ spatial modes that discriminates the subspace with zero or one
photon from the rest\togli{ (but which does not distinguishes among
 the zero photon subspace and the one photon subspace)}. If he finds
that the modes jointly contain more than one photon, he knows that
Alice is trying to violate the data privacy, and stops the
communication. If, instead, he finds that Alice is sending no more
than one photon per query, he can be sure that she is recovering no
more than one bit per transaction.

Unfortunately, the above measurement is practically unfeasible. An
alternative solution which is feasible, although less efficient, is
the following.  
After Alice
has sent her first photon into his lab, 
Bob blocks the access to the database and  partitions it into $X$ equal parts $P_1, P_2, \cdots, P_X$
containing $N/X$ random entries each.
He then communicates to Alice the composition of the partitions asking to 
reveal $\log_2 X$ bits on her query to 
indicate 
 which of the $P_\ell$'s contains the database entry she is interested in
 (the fact that Alice has to reveal some bits  should not be seen as a breach of the user privacy, since this is a (small) fixed quantity which is independent on the database size). 
Bob now can perform a local photodetection on each
of the modes of the $X-1$ partitions which according to Alice do not contain
  the message she is looking for.
  If he finds any photons there, he knows for sure that Alice has
  cheated and stops the communication. If he does not, he cannot
  conclude that Alice has cheated and allows her to complete her query
  sending the second photon, for which the above procedure is
  repeated.

\begin{figure}[t!!]
\centering
\includegraphics[width=4.2cm]{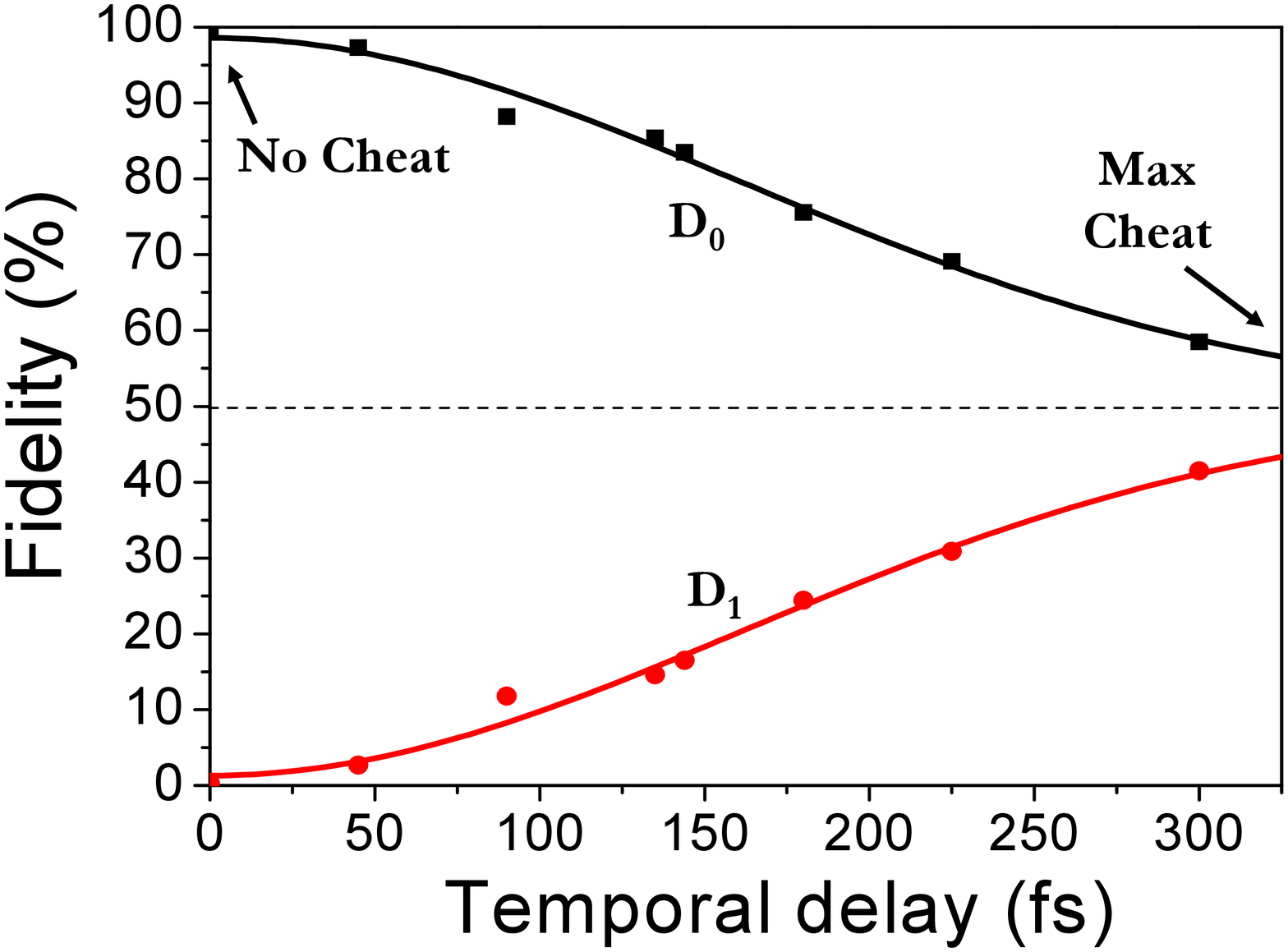}\hspace{.3cm}
\includegraphics[width=4cm]{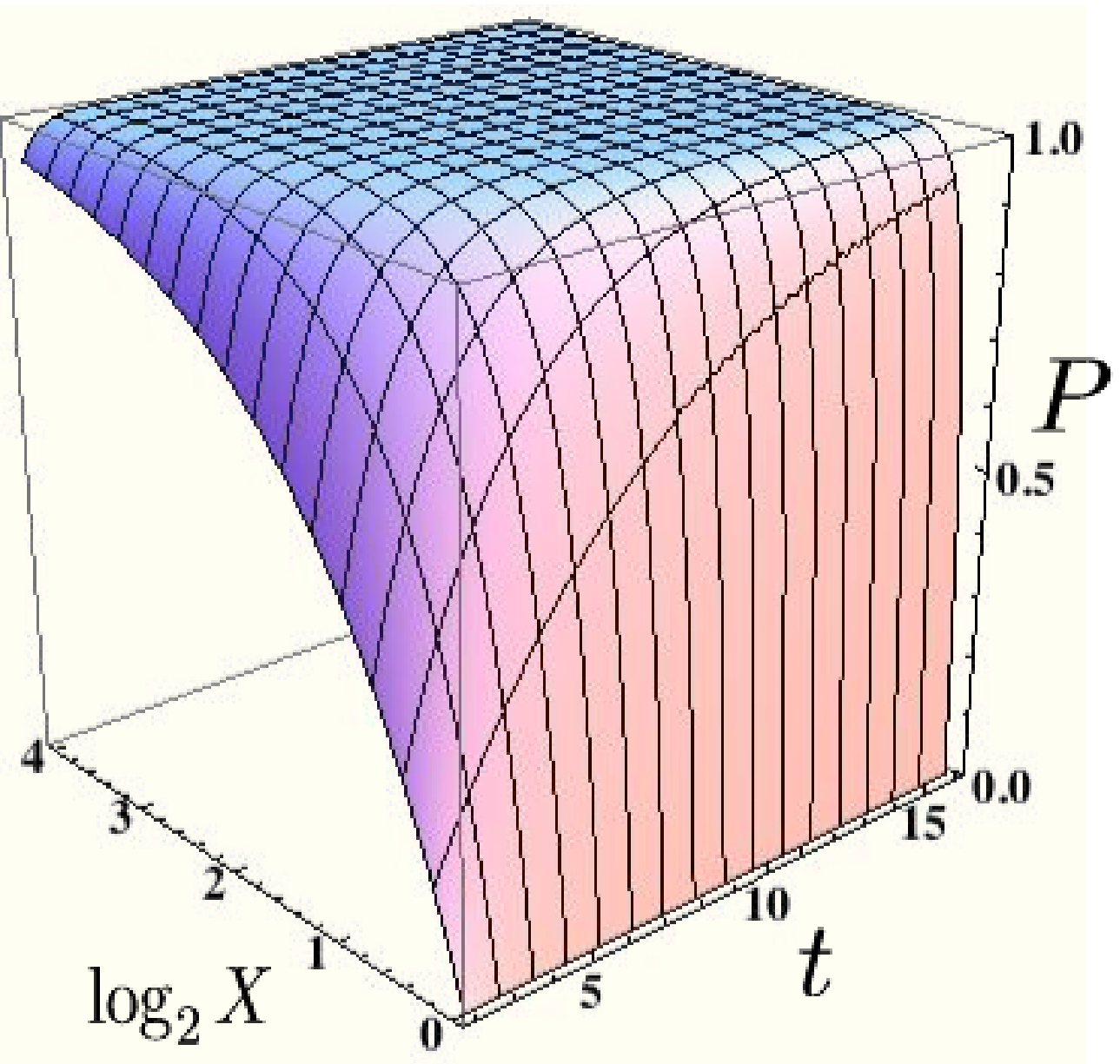}
\caption{Left: Experimental fidelity values (dots) for different time
  delays introduced in the interferometer, which simulate the effects
  of different cheating attacks by Bob (large values of the temporal
  delay correspond to larger disturbances, i.e.~to a larger
  information capture by Bob). The upper curve is the fit for the
  probability that Alice's ``don't know'' detector D$_0$ fires during
  the honesty test, the lower curve refers to probability that Alice's
  ``cheat'' detector D$_1$ fires. The fit function is Gaussian due to the spectral and temporal profile of the single photon state.
  Right: Theoretical curve
  representing the data privacy $P= 1 - (1/X)^{t-1}$ as a function of the bits $\log_2X$ Alice reveals to Bob and of the photons $t$  she uses to cheat.}\label{figure2}
\end{figure}

As in the case of  user privacy, the data privacy is thus enforced by 
means of  a probabilistic, non conclusive honesty test. 
In particular  there is a  tradeoff:
 the more bits Alice
reveals on her query, the higher is the probability that Bob will be able to find out if she is cheating. 
For instance, consider the case in which  Alice tries to recover some extra bits from the database by
 sending $t\geqslant1$  photons per transmitted signal. 
Assuming random encodings, the probability that all of them  will be found in the same subset of the database partition can be estimated as $X (1/X)^t = (1/X)^{t-1}$. 
This is the only case in which 
Alice can safely  pass Bob's honesty test. In all remaining cases at least one of the $t$ photons   will belong to one of the subsets on which  Bob performs his photodetections. Alice's probability of being
caught is thus equal to $P= 1 - (1/X)^{t-1}$, which increases both with the number $(t-1)$ of cheating photons  and with  the number   $\log_2 X$  of   bits she reveals to Bob
-- see Fig.~\ref{figure2}. The gating is also fundamental: Bob must open the access to the
database only during the transit time of Alice's photons, prompted by
a trigger signal. Otherwise, she can cheat sending photons at other
times. Similar expedients are usually adopted in plug-\&-play
cryptographic schemes to avoid Trojan horse attacks \cite{gisin3}. These parts of the protocol are important only if data privacy
is an issue. As done in the experiment below, it can be omitted when
only user privacy is important.

\textbf{Experimental results.} In order to perform a
proof-of-principle experiment, we have to show that Alice can recover
the value of each database element, and that she can detect Bob's
cheats. The single photon is created by starting from a biphoton
generated through spontaneous parametric downconversion and using one
of the two component photons as a trigger. A sequence of half-wave
plates and polarizing beam splitters allows Alice to choose the mode
$j$ (i.e.~the database element) she wants to access with her ${H}$
polarized photon -- see Fig.~\ref{f:schemablocchi}(b). In the
experiment we employed $N=3$ modes\togli{, where Bob's database is
  encoded on the presence or absence of polarization rotators}. A
standard polarization analysis setup and single photon detectors
implement the reading process of Fig.~\ref{f:schemablocchi}(d)
performed by Alice.  In Table~\ref{table1}-(a) we report the experimental
results for the preparation and measurement of each query
$|{P}_j\rangle$ ($j=1,\cdots,3$), giving the outcome fidelity for each
element in the database.

\begin{table}[b]
{\footnotesize
\begin{center}
%\textbf{Reading process}\\[2pt]
\begin{tabular}{|c||c|c|| c|c|}
  \hline $\text{$j$}$ & $\;\;A_j\;\;$ & Fidelity &
  $\;\;A_j\;\;$ & Fidelity\\ \hline\hline $1$ & $0$ &
  $(99.84\pm0.04)\%$ & $1$ & $(99.99\pm0.01)\%$\\ \hline $2$ & $0$ &
  $(99.81\pm0.04)\%$ & $1$ & $(99.72\pm0.05)\%$\\\hline
  $3$ & $0$ & $(99.99\pm0.01)\%$ & $1$ & $(99.99\pm0.01)\%$\\
  \hline\hline
\end{tabular}
\begin{tabular}{|c||c||c|c||c|c|}
\hline
$$ & $\;\;A_j\;\;$ & $D_0\;th.$ & $D_0\;exp.$ & $D_1\;th.$ & $D_1\;exp.$\\ \hline\hline
$No \; cheat$ & $0$ & $1$ & $(99.3\pm0.2)\%$ & $0$ &
$(0.7\pm0.2)\%$\\ \hline $No \; cheat$ & $1$ & $1$ &
$(99.4\pm0.2)\%$ & $0$ & $(0.6\pm0.2)\%$\\\hline\hline $Cheat$ &
$0$ & $0.5$ & $(45.0\pm0.1)\%$ & $0.5$ & $(55.0 \pm
0.1)\%$\\\hline
$Cheat$ & $1$ & $0.5$ & $(45.4\pm 0.1)\%$ & $0.5$ & $(54.6 \pm0.1)\%$\\
\hline\hline
\end{tabular}
\\
\end{center}
}
\par
%\vspace{0.2cm}
%\noindent
%TABLE I.
\caption{\textbf{(a)} Experimental values of the fidelity of
  Alice's measurement of each of the three elements of Bob's
  database. The measurement is performed by sending queries of the
  form $|{P}_j\rangle$, and measuring the output polarization---
  see Fig.~\ref{f:schemablocchi}(d). \textbf{(b)} Comparison between theoretical (\textit{th.})
  and experimental (\textit{exp.}) fidelities of Alice's honesty test of
  Fig.~\ref{f:schemablocchi}(c). The discrepancy with theory is due to unbalancement of the interferometer and slight misalignment.}\label{table1}
\end{table}

The characterization of the honesty test follows. Alice must be able
to move the interferometer of Fig.~\ref{f:schemablocchi}(c) to the
mode $j$ corresponding to the question she wants to ask.  We have
implemented this using a Jamin-Lebedeff interferometer, which is quite
compact, easy movable, and leads to a high phase stability
\cite{OBri03}. In the first part of Table~\ref{table1}-(b) we
characterize Alice's honesty test when Bob is not cheating.
To cheat, Bob may introduce a beam splitter in each mode and place a
detector at the beam splitter output port. When he detects a photon in
a mode $j$, he recreates a photon there. To simulate this cheating
attack, we introduced a variable time delay in each mode. A delay
larger than the photon's coherence length simulates a
``measure-and-reprepare'' cheat (i.e.~zero beam splitter
transmissivity). Shorter delays simulate a milder cheat (i.e.~nonzero
beam splitter transmissivities). This was implemented by inserting
quartz plates of varying thickness in Bob's arm of the interferometer. In
Table~\ref{table1}-(b) and in Fig.~\ref{figure2} Alice's honesty test is
characterized also in the presence of cheating. 

\textbf{Time-slot implementation.} We now describe a different
implementation of the scheme, based on~\cite{gisin1,gisin2}. To each
database element $j$ we associate a unique time slot in an optical
fiber: Alice places her query photon in the $j$th slot (i.e.~the state
$\ket{P_j}$) if she wants to access $A_j$. Bob's database is encoded
into a time dependent polarization rotator: in the $j$th time slot the
polarization is rotated only if $A_j=1$.  To create the superposed
$\ket{S_j}$ query, Alice places her photon in a superposition of two
time slots~\cite{gisin1}. This is achieved by sending it through a
$50\%$ beam splitter, at the two outputs of which she places a long
and a short fiber. The length difference of the fibers corresponds to
a delay proportional to $j$. The signals from the two fibers are then
joined into a single fiber through an optical switch~\cite{gisin1}.
The same device (used in reverse) is used as cheat test on the
superposed signal returning from Bob: the optical switch sends the
first pulse through the long fiber and the second through the short
fiber, so that they interfere at the beam splitter.  The photon then
exits at one of the two ``cheat'' or ``don't know'' ports of the beam
splitter. It is simple to see that this implementation is conceptually
equivalent to the previous one, but it is more suited to the case in
which Alice and Bob are far apart, as this procedure has been tested
experimentally with interferometers of many Km in
length~\cite{gisin2,gisin3}. Our protocol can be easily scaled up considerably since the resources scale only linearly with the number of database elements. The number of database elements is ultimately limited only by the time-dependent noise the photons encounter along the fiber.

\end{document}